\documentclass[10pt]{article}
\usepackage[top=0.85in,left=1.25in,footskip=0.75in]{geometry}

\usepackage{amsmath,amssymb}

\usepackage{changepage}

\usepackage[utf8x]{inputenc}

\usepackage{textcomp,marvosym}

\usepackage{cite}

\usepackage{nameref,hyperref}

\usepackage{microtype}
\DisableLigatures[f]{encoding = *, family = * }

\usepackage[table]{xcolor}

\usepackage{array}

\newcolumntype{+}{!{\vrule width 2pt}}

\newlength\savedwidth

\usepackage{setspace} 
\usepackage[aboveskip=1pt,labelfont=bf,labelsep=period,justification=raggedright,singlelinecheck=off]{caption}

\makeatletter
\renewcommand{\@biblabel}[1]{\quad#1.}
\makeatother

\usepackage{lastpage,fancyhdr,graphicx}
\usepackage{epstopdf}
\fancyheadoffset[L]{2.25in}
\fancyfootoffset[L]{2.25in}
\usepackage{float}

\begin{document}
\vspace*{0.2in}

\begin{flushleft}
{\Large
\textbf\newline{Emergence of linguistic conventions in multi-agent reinforcement learning} % Please use "sentence case" for title and headings (capitalize only the first word in a title (or heading), the first word in a subtitle (or subheading), and any proper nouns).
}
\newline
\\
Dorota Lipowska\textsuperscript{1},
Adam Lipowski\textsuperscript{2*}
\\
\bigskip
\textbf{1} Faculty of Modern Languages and Literature, Adam Mickiewicz University, Pozna\'{n}, Poland
\\
\textbf{2} Faculty of Physics, Adam Mickiewicz University, Pozna\'{n}, Poland
\\
\bigskip

% Use the asterisk to denote corresponding authorship and provide email address in note below.
Corresponding author:\\
* lipowski@amu.edu.pl

\end{flushleft}
\section*{Abstract}
Recently, emergence of signaling conventions, among which language is a prime example, draws a considerable interdisciplinary interest ranging from game theory, to robotics to evolutionary linguistics. Such a wide spectrum of research is based on much different assumptions and methodologies, but complexity of the problem precludes formulation of a unifying and commonly accepted explanation. We examine formation of signaling conventions in a framework of a multi-agent reinforcement learning model. When the network of interactions between agents is a complete graph or a sufficiently dense random graph, a global consensus is typically reached with the emerging language being a nearly unique object-word mapping or containing some synonyms and homonyms. On finite-dimensional lattices, the model gets trapped in disordered configurations with a local consensus only. Such a trapping can be avoided by introducing a population renewal, which in the presence of superlinear reinforcement restores an ordinary surface-tension driven coarsening and considerably enhances formation of efficient signaling.
%\linenumbers

% Use "Eq" instead of "Equation" for equation citations.
%%%%%%%%%%%%%%%%%%%%%%%%%%%%%%%%%%%%%%
%%%%%%%%%%%%%%%%%%%%%%%%%%%%%%%%%%%%%%
\section*{Introduction}
Functioning of societies is to a large extent regulated by various norms and conventions shared by its members~\cite{lewis}. In some cases, these rules are centrally imposed or coordinated, e.g., a dress code in a company or the side of the road that one should drive on. But some conventions, such as the color of cloth that we wear in grief or greeting our friends with a handshake, appeared more spontaneously. 

Perhaps the most important convention of this kind, which emerged in the absence of any explicit, centralized coordination, is language. Human language provides a highly efficient communication system acquired by individuals  in cultural interactions. Some researchers try to explain how such a system could have appeared and evolved using evolutionary game theory~\cite{nowak,komarova}, evolutionary linguistics~\cite{oliphant} or cognitive science~\cite{barr}. A promising approach considers language as a signaling system, which emerged via a reinforcement learning process. Such a framework originates from Lewis signaling game~\cite{lewis}. In the simplest version, there are two players and a fixed number of signals. The speaker  sends a signal (which is to correspond to the state of the world) and the hearer interprets the signal (i.e., takes some action). If he does it correctly, both players receive  some payoff, which might influence their further actions. The model can actually be considered as a certain urn model~\cite{ep23,pm2007}. Some mathematical subtleties concerning, e.g., the convergence of the above scheme, were analyzed by Skyrms~\cite{skyrms} and Beggs~\cite{beggs}, while an adaptation focusing on language evolution was proposed by Lenaerts {\em et~al.}~\cite{lenaerts}. Attempts to compare several related approaches where learned signaling  might emerge were also made~\cite{kirby}. 

In all these studies, the examined number of agents was rather small ($\lesssim 30$), however, one should note that the reinforcement learning leads to nontrivial results already in two-agent  models~\cite{lewis,skyrms,barrett}.  Nonetheless, having in mind the evolution of language, much larger populations of agents should certainly be examined. In such a case, for a population of agents, one has to specify the network of their interactions. While for a small group, a complete graph, where each agent interacts with each of the others, seems the most natural topology, for larger groups of agents, some other structures such as planar or heterogeneous graphs can also be relevant (e.g., when studying the emergence of a linguistic coherence in large-scale communities such as  a city population or a nation). 
The emergence of language is often modeled as a process of reaching an agreement (consensus) about linguistic forms used in a population. Opinion formation or ferromagnetism are also manifestations of such an agreement dynamics. For such processes, the structure of the network usually plays an important role, determining whether the consensus will be reached at all, and affecting the way it could be reached~\cite{rmp,baronchelli2018,lip2017}. Networks examined in the present paper (Cartesian lattices, complete graphs, random graphs) are only mathematically and computationally appealing idealizations of real networks. Certainly, placing our models on more realistic networks, which take into account a node-distribution heterogeneity, directionality, small-worlds, modular structure or assortativity \cite{kaski}, would be desirable.

A model that is often examined in the context of language emergence  is the Naming Game~\cite{steels}. Due to its computational simplicity, the Naming Game allows for analytical as well as numerical approaches, and global aspects of its dynamics are now relatively well understood~\cite{baron}. In particular, it is found that typically in the Naming Game, a consensus emerges and reaching such a state resembles the coarsening in the Ising model. The  similarity is not accidental because due to the presence of a surface tension~\cite{ng-surface-tension}, both models operate with the so-called curvature-driven dynamics~\cite{bray}.  Let us notice that the coarsening dynamics of the Naming Game, which gradually eliminates certain languages and eventually leads to a global consensus, can be found very appealing in some linguistic contexts. There are even some indications that the curvature-driven dynamics may underlie such  linguistic processes as, e.g., an evolution of dialects~\cite{human-dialects}.  The simplicity of the Naming Game implies, however,  simplicity of an emerging language, and in many of its versions agents negotiate the name of just a single object. On the other hand, for models that have a potential to generate more complex languages,  global aspects of their dynamics  are rather poorely understood. Such models could incorporate agents, which, using the reinforcement learning, would try to establish a language reflecting their multi-object and multi-agent world. An objective of the present paper is to specify whether and how an efficient communication might emerge in such a system.

%%%%%%%%%%%%%%%%%%%%%%%%%%%%%%%%%%%%%%
%%%%%%%%%%%%%%%%%%%%%%%%%%%%%%%%%%%%%%
\section*{Methods}

%%%%%%%%%%%%%%%%%%%%%%%%%%%%%%%%
\subsection*{Reinforcement learning via urn model}

The basic building block of our model is a P\'olya urn model. In the simplest version of this model, a ball is drawn randomly from an urn with black and white balls~\cite{ep23,pm2007}. Then the ball is put back into the urn along with an extra ball of the same color (reinforcement), and the process is repeated {\it ad infinitum}. In this scheme, the probability to select a ball of a given color is proportional to the number of such balls in the urn. We can also consider a generalized version of this model with the selection probability proportional to the number of balls raised to a certain power~$\alpha$~\cite{drinea2002}. In this case, the behavior of the model strongly depends on~$\alpha$. For $\alpha<1$, the model converges toward an equal number of balls of each color, but for $\alpha>1$, a monopolistic solution appears with the urn dominated by one color. The monopolistic solution is in fact a simple manifestation of a spontaneous symmetry breaking, the phenomenon of much interest in statistical mechanics or particle physics.  The basic P\'olya urn model is equivalent to the $\alpha=1$ case,  thus determining the transition between these two different regimes.

Our intention is to study a multi-agent model of a signaling game with communicating agents as interacting urns. In the simplest (single-object) version, agents engage in  pairwise interactions to negotiate the word to be associated with an object. After a weighted selection of a word, the speaker and the hearer increase its weights (reinforcement learning), which affects subsequent selections. It seems plausible that in the $\alpha>1$ regime, a monopolistic solution would emerge with agents almost always selecting the same word. There is, however, a number of questions, which one can ask concerning such a linguistic consensus. For example, is it a global consensus, where the entire population of agents uses the same word, or rather a local one corresponding to a certain multi-word solution. Most likely the answer will depend on the topology of interactions between agents, e.g., networks of long-range connectivity should favour the global consensus.  Furthermore, agents may be involved in more complicated interactions, e.g., negotiating simultaneously the names for several objects (multi-object version). In that case they need some recognition mechanism, and the resulting language is likely to be more complex. 

It is difficult to advocate that in the linguistic contexts, $\alpha>1$ should be used. In economy, the emergence of a monopoly is sometimes associated with a certain positive superlinear feedback known as Metcalfe's Law~\cite{shapiro}. For example, in social networks, the greater the number of users with a certain service, the more valuable the service becomes to the community, and hence its total value is likely to increase quadratically ($\alpha=2$) with the number of its users. 
One might expect that a similar superlinear feedback appears during language formation processes. Most of the results presented in our paper are for $\alpha=2$; some of our results demonstrate that the behaviour of the model is qualitatively similar as long as $\alpha>1$. For $\alpha=1$ the convergence toward consensus is typically much slower and in some cases the model does not evolve toward consensus at all.

%%%%%%%%%%%%%%%%%%%%%%%%%%%%%%%%
\subsection*{Single-object version} 

\begin{figure}[!h]
 \centering
\includegraphics[width=\columnwidth]{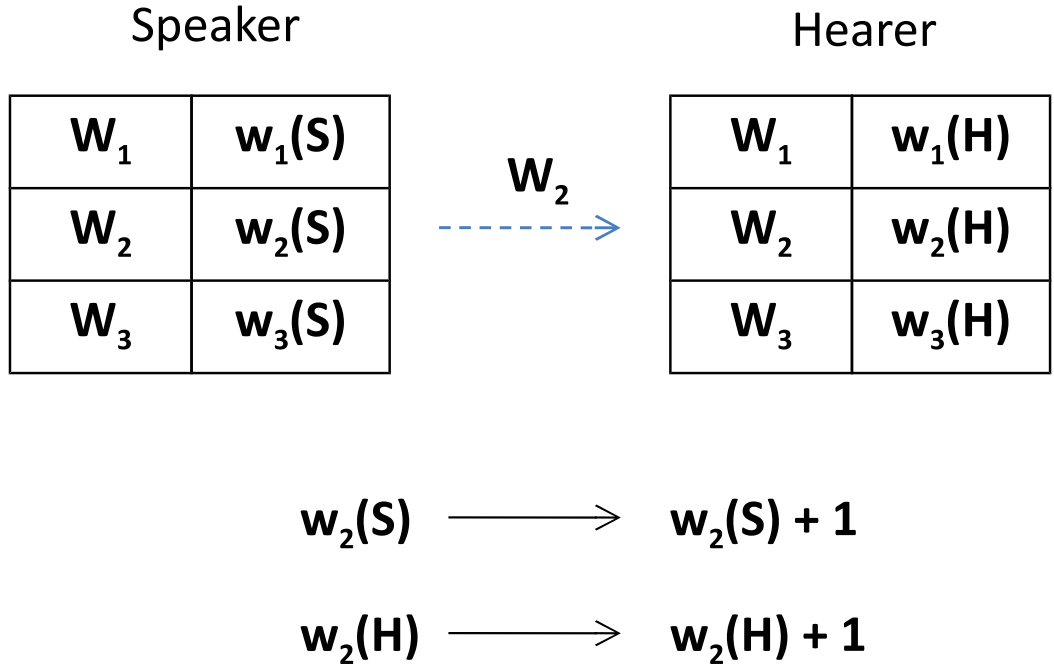}
\caption{  {\bf An elementary step of a single-object version of the model ($\boldsymbol{N_w=3}$). }
Using the probabilities defined in Eq~(\ref{eq1}), the speaker selects one of its words (here: $W_2$). Next both the speaker and the hearer increase their weights of the selected word by~1.}
\label{Fig1}
\end{figure}

In the simplest version of our model, we have a population of $N$~agents, which try to establish a name for a given object. Each agent~$A$ has an inventory of the same $N_w$ words $W_i$ with their corresponding weights $w_i(A)$ ($i=1,2,\ldots, N_w$; initially all $w_i(A)=1$).  In an elementary step, a randomly selected agent (the speaker) interacts with  one of its randomly selected neighbors (the hearer) communicating a word. The probability that the speaker~$A$ will select  the $i$-th word depends on its weight and is given as
\begin{equation}
s_i(A)=w_i^{\alpha}(A)/\sum_{k=1}^{N_w} w_k^{\alpha}(A). 
\label{eq1}
\end{equation}
After the interaction, both the speaker and the hearer increase their weights of the communicated word by~1. Such an elementary step of our model is illustrated in Fig~\ref{Fig1}. In our simulations, a unit of time (t=1) comprises $N$~elementary steps (i.e., in a unit of time, each agent is on average selected once as a speaker). 

%%%%%%%%%%%%%%%%%%%%%%%%%%%%%%%%%%%%%%%%%%%%%%%
%%%%%%%%%%%%%%%%%%%%%%%%%%%%%%%%

%%%%%%%%%%%%%%%%%%%%%%%%%%%%%%%%
\subsection*{Multi-object version} 
\begin{figure}[!h]
 \centering
\includegraphics[width=\columnwidth]{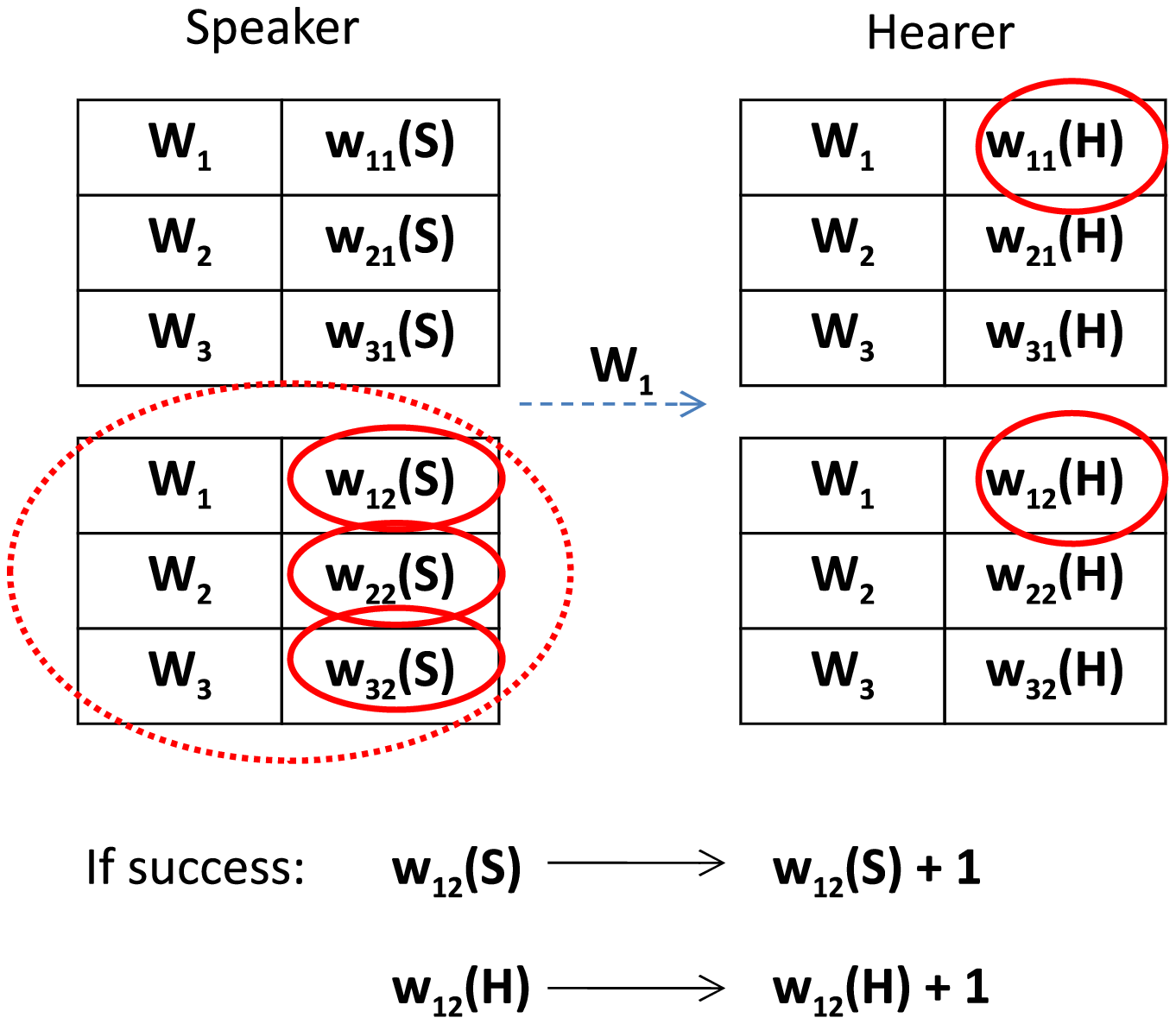}
\caption{ {\bf An elementary step of a multi-object version of the model 
($\boldsymbol {N_w=3}$, $\boldsymbol {N_o=2}$). }
 With a uniform probability $1/N_o$, the speaker chooses an object (the corresponding section of the inventory is encircled by the dotted line). Using the relevant weights (in solid circles), the speaker calculates the probabilities defined in Eq~(\ref{eq2}) and selects one of its words (here: $W_1$). Next the hearer tries to guess the object the speaker is talking about by calculating the probabilites~(\ref{eq3}) based on its weights of the communicated word (in circles). When the hearer's guess is correct, both agents increase their corresponding weights by~1.}
\label{Fig2}
\end{figure}

We also examine a more general version of our model, in which agents try to establish names for a set of $N_o$~objects.  Their inventories are more complex now as they contain the same set of $N_w$ words $W_i$ (coupled with their respective weights) for each object. In other words, each inventory consists now of $N_o$ copies of inventories from the single-object version and thus each agent~$A$ has $N_wN_o$ weights $w_{i,j}(A)$, where $i=1,\ldots, N_w$  and  $j=1,\ldots, N_o$.  First, a randomly selected speaker chooses an object with a uniform probability $1/N_o$. Then the speaker selects the word to be communicated taking into account the weights associated with the words for the chosen object. By analogy with Eq~(\ref{eq1}), the probability that agent~$A$  will select the $i$-th word for the $j$-th object equals 
\begin{equation}
s_{i,j}(A)=w_{i,j}^{\alpha}(A)/\sum_{k=1}^{N_w} w_{k,j}^{\alpha}(A).
\label{eq2}
\end{equation}
Next, the role of the hearer~($H$) is to assign an object to the communicated word. This word, say $W_i$, appears in the hearer's inventory  $N_o$ times with weights $w_{i,j}(H)$, where $j$ denotes the object. The hearer uses these weights to guess which object the speaker is talking about. Hence, the hearer recognizes the $j$-th object as that communicated by the $i$-th word with probability 
\begin{equation}
r_{i,j}(H)=w_{i,j}^{\alpha}(H)/\sum_{k=1}^{N_o} w_{i,k}^{\alpha}(H).
\label{eq3}
\end{equation}
 Provided that the object recognized by the hearer is the same as that chosen by the speaker, both agents increase the corresponding weights by~1. An elementary step of this version of the model is  illustrated in Fig~\ref{Fig2}.

%%%%%%%%%%%%%%%%%%%%%%%%%%%%%%%%%%%%%%%%%%%%%%%
%%%%%%%%%%%%%%%%%%%%%%%%%%%%%%%%

The above specified rules are consequences of a number of simplifying assumptions and certainly more realistic versions might be considered. For example, one might assume that the words in agents' inventories are not necessarily identical and agents could learn new words from each other. Most likely such a change would require a more sophisticated recognition mechanism and perhaps a notion of a distance between words would have to be used.  Further analysis of such a version, although it seems more realistic and potentially interesting, is left for the future.

%%%%%%%%%%%%%%%%%%%%%%%%%%%%%%%%
\subsection*{Population renewal} 

We also introduce a simple modification of our model (both in its single and multi- object versions), which takes into account a population renewal. The modification seems to be plausible, especially for modeling a formation of a communication system in a population of humans. In such population, when considered at a timescale of, say, hundreds of years, we should take into account a generational turnover (and possibly migrations~\cite{lipmigracje}). A child learns the language of its parents but it might also acquire a (possibly different) language of its  neighbors. Certainly, for a young person this is more likely to happen than for an adult. Let us notice that in urn models, due to the accumulation of weights after a large number of iterations, it is almost impossible to shift their balance (i.e., change the language). To allow for such a shift,  we introduce a population renewal:  With (usually small) probability~$p$, the agent selected to be a speaker is replaced with a new agent (with all weights equal to~1), while with probability \mbox{$1-p$}, the speaker acts as previously defined. 

%%%%%%%%%%%%%%%%%%%%%%%%%%%%%%%%%%%%
%%%%%%%%%%%%%%%%%%%%%%%%%%%%%%%%%%%%
% Results and Discussion can be combined.
\section*{Results}

%%%%%%%%%%%%%%%%%%%%%%%%%%%%%%%%
\subsection*{Single-object version} \label{single-object}

\begin{figure}[!h]
\includegraphics[width=\columnwidth]{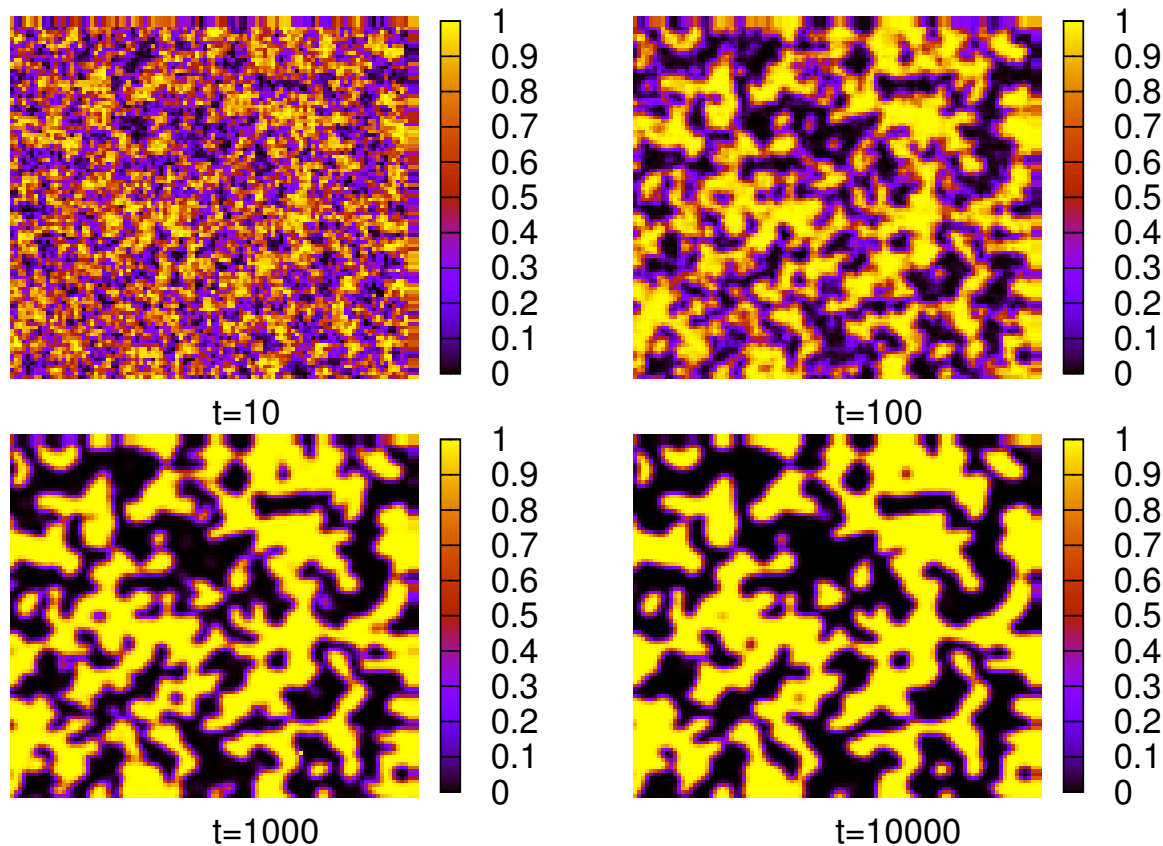}
\caption{ {\bf Spatial distribution of $\boldsymbol{s_1}$,  the probability that an agent will select  word $\boldsymbol{W_1}$ (Eq~(\ref{eq1})).}
Results for a single-object model on a square lattice with $N=10^2\cdot 10^2=10^4$, $\alpha=2$, $N_w=2$. The dynamics traps the model in a disordered state (the configurations for, e.g., $t=10^3$ and $t=10^4$ differ only slightly). Since $s_1$ is generally close to unity or to zero, it means that almost every agent developed a strong preference toward one of the words.}
\label{Fig3}
\end{figure}

We analysed the behavior of our model for several interaction networks, namely Cartesian lattices, complete graphs and random graphs. The results obtained indicate that on Cartesian lattices the model gets stucked in a disordered structure, where consensus is only local (Fig~\ref{Fig3}).  Most of the agents reach a monopolistic regime and communicate with their neighbors with the same words. There is only a small fraction of interfacial agents, which persist in a more symmetric state.   For $N_w=2$, it is tempting to confront our results with some other statistical-mechanics models. In particular, the snapshot configurations in Fig~\ref{Fig3} suggest  that initally our model coarsens, similarly to the Naming Game and low-temperature Ising models. However,  contrary to these models, the evolution of our model gets trapped in a disordered state much before reaching the uniform (mono-word) state. 

%%%%%%%%%%%%%%%%%%%%%%%%%%%%%%%%%%%%%%%%%%%%%%%
%%%%%%%%%%%%%%%%%%%%%%%%%%%%%%%

To examine in more detail the behavior of the model, we calculated for $N_w=2$ the quantities $m_G$ and $m_L$ defined as
\begin{align}
m_G&=|\frac{1}{N}\sum_A (s_1(A)-s_2(A))|, \nonumber \\ 
m_L&=\frac{1}{N}\sum_A |s_1(A)-s_2(A)|
\label{qm}
\end{align}
where summation is over all agents $A$ in our model.

The quantities $m_G$ and $m_L$ allow us to examine the global ($m_G$) and local ($m_L$) symmetries of the model. We do not present the results for $\alpha<1$, in which case the model remains symmetric (for each agent, $s_1=s_2=1/2$), and consequently, $m_G=m_L=0$. For $\alpha>1$, however,  the asymmetry in an agent's inventory implies that $s_1\neq s_2$ and thus $m_L>0$. When the system is disordered, as in Fig~\ref{Fig3}, then there is no global preference toward any of the two words and $m_G=0$.

\begin{figure}[!h]
\begin{center}
\includegraphics[width=13cm]{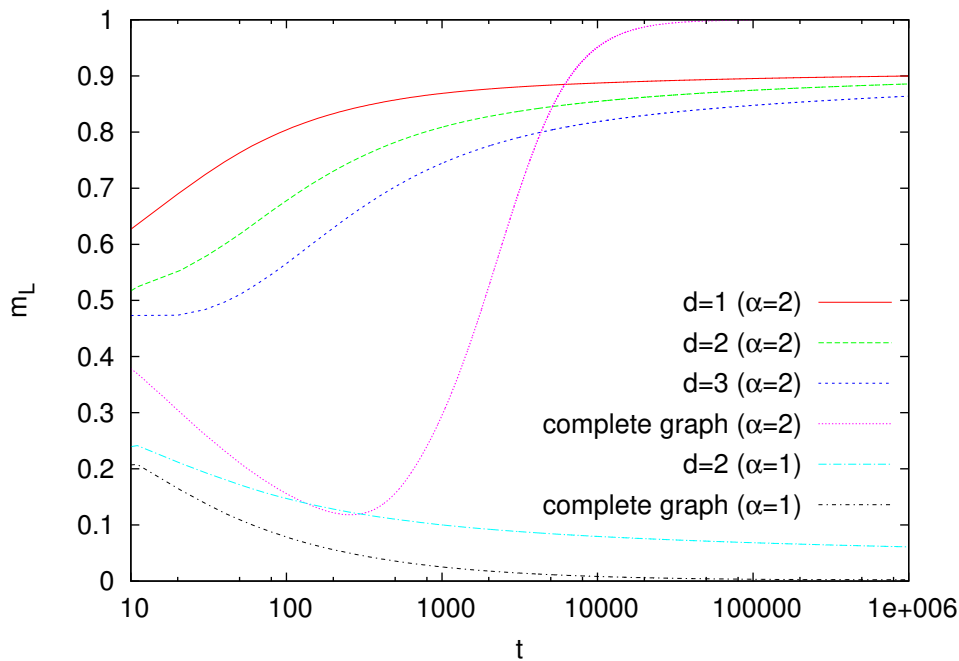}
\end{center}
\caption{{\bf Time dependence of $\boldsymbol {m_L}$.} 
Results for a single-object model with $N_w=2$ on complete graphs ($N=10^5$) and  Cartesian lattices with $d=1$ ($N=10^5$), $d=2$ \mbox{($N=300^2=9\cdot 10^4$)}, and $d=3$ ($N=50^3=125\cdot 10^3$). In the case of the ordinary reinforcement ($\alpha=1$), none of the words is even locally preferred on a complete graph (since $m_L\rightarrow 0$), and only a small asymmetry is seen for a square lattice. The results presented  (also in the following figures) are averages over 20 independent runs. Statistical errors are typically smaller than plotting symbols and are omitted.}
\label{Fig4}
\end{figure}

\begin{figure}[!h]
\begin{center}
\includegraphics[width=13cm]{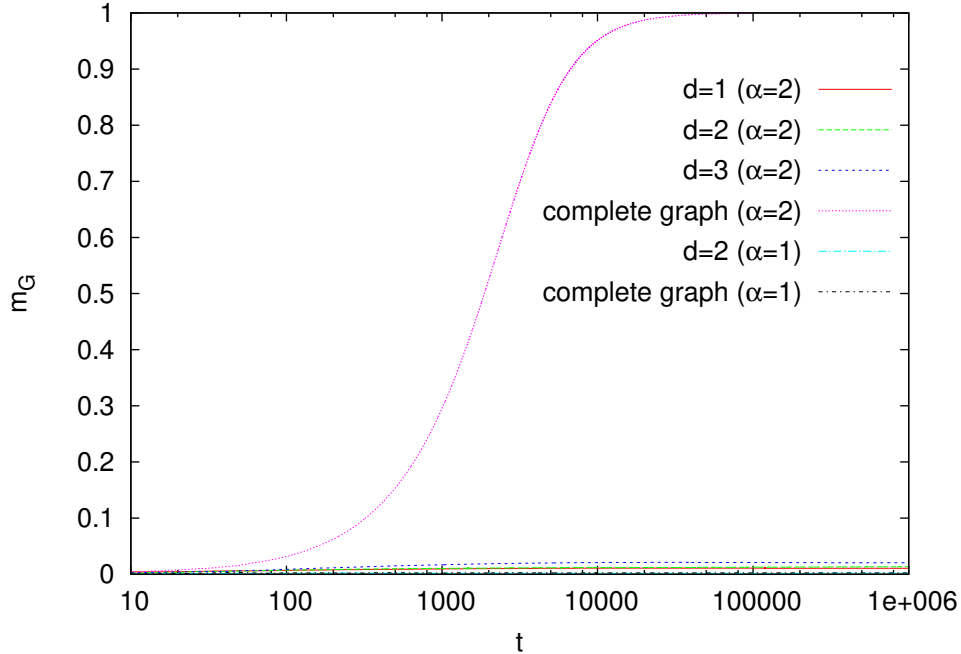}
\end{center}
\caption{ {\bf Time dependence of $\boldsymbol {m_G}$.}
Results for a complete graph and Cartesian lattices with the same simulation parameters as in Fig~\ref{Fig4}. Only for a complete graph and $\alpha=2$, a global symmetry gets broken and one word dominates in the entire population of agents.}
\label{Fig5}
\end{figure}

Numerical results for $\alpha=2$ support such an analysis. In two dimensions, relatively large $m_L$ (Fig~\ref{Fig4}) indicates that most of the agents operate in a monopolistic regime. Moreover, $m_G$  remains close to zero, which confirms the disordered nature of the regime (Fig~\ref{Fig5}).
Results for the one-  and three-dimensional Cartesian lattices show a similar behavior.  Much different behavior is seen, however, for a complete graph, where each agent interacts with every other one. In this case, $m_G$ quickly reaches unity, which indicates that basically all agents communicate using the same word. It means that on the complete graph not only the local symmetry is broken ($m_L>0$) but also the global one ($m_G>0$). For $\alpha=1$, simulations for both a square lattice and a complete graph show that $m_L$ is small (and decreases in time) and thus even the local symmetry is preserved (Fig~\ref{Fig4}).

%%%%%%%%%%%%%%%%%%%%%%%%%%%%%%%%%%%%%%%%%%%%%%%
%%%%%%%%%%%%%%%%%%%%%%%%%%%%%%%

%%%%%%%%%%%%%%%%%%%%%%%%%%%%%%%%%%%%%%%%%%%%%%%
%%%%%%%%%%%%%%%%%%%%%%%%%%%%%%%

\begin{figure}[!h]
\begin{center}
\includegraphics[width=13cm]{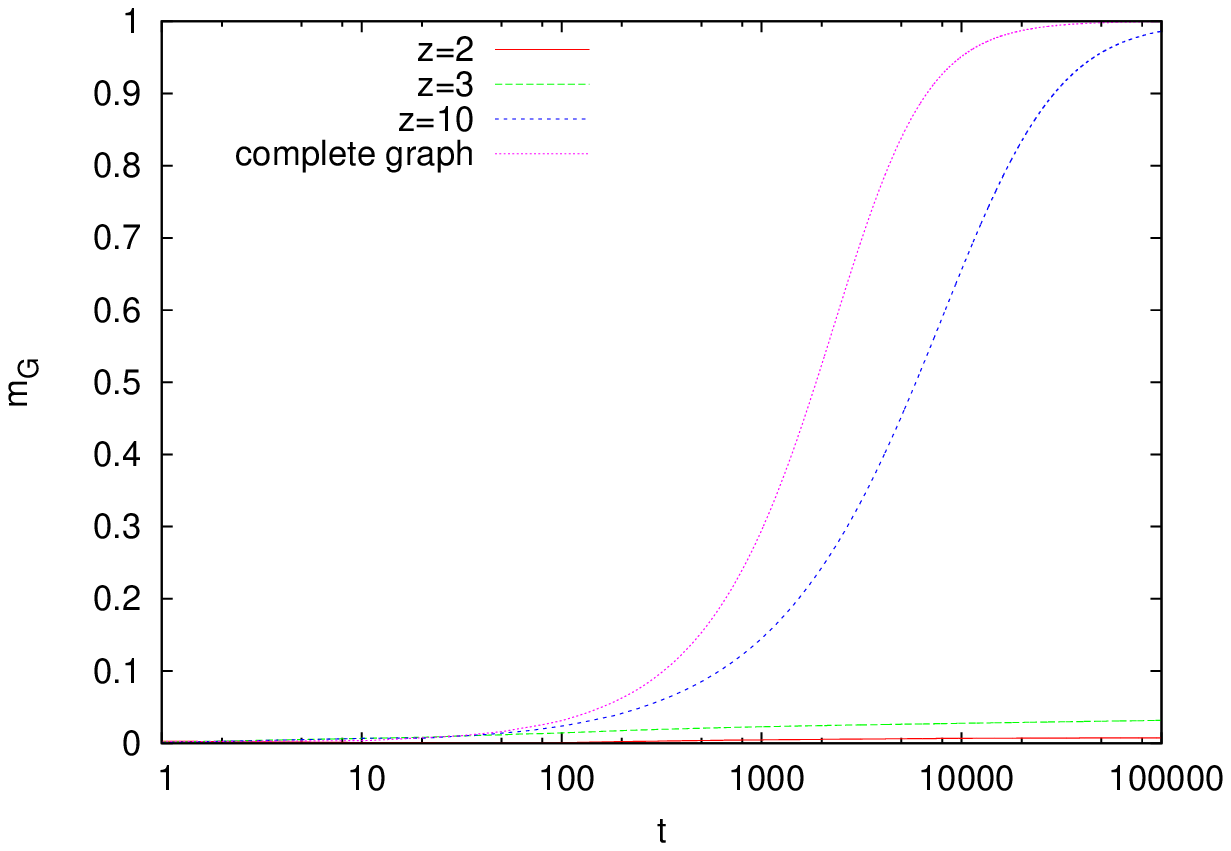}
\end{center}
\caption{ {\bf Time dependence of $\boldsymbol {m_G}$ (random graphs).}
Results for random graphs of an average node degree~$z$ and a complete graph ($\alpha=2$, $N_w=2$, $N=10^5$). Only for sufficiently large~$z$, the behavior on the random graphs is similar to that on the complete graph. Averaging over 20 runs includes generation  of independent graphs.}
\label{Fig6}
\end{figure}

Having in mind modeling the emergence of communicative consensus in a population of agents, it is desirable to examine the behavior of our model also on heterogeneous networks. The simplest ones are perhaps random graphs. We examined our model on Erd\"os-R\'enyi random graphs~\cite{erdos,newman} of an average node degree~$z$. For large~$z$ ($z=10$), the model behaves similarly as on a complete graph and quickly reaches a global consensus about the communicated word (Fig~\ref{Fig6}). For smaller~$z$ ($z=2, 3$), the model remains trapped in  a disordered phase, where consensus is reached only locally  (similarly as on a  square lattice).

%%%%%%%%%%%%%%%%%%%%%%%%%%%%%%%%%%%%%%%%%%%%%%%
%%%%%%%%%%%%%%%%%%%%%%%%%%%%%%%

Let us recall that for random graphs,  $z=1$ marks the percolation transition~\cite{newman}. To study the formation of a global consensus, one needs to consider only $z>1$, since for $z<1$ the graph decomposes into separate components. Random graphs for finite~$z$ are tree-like hence they are effectively infinite-dimensional. One might thus expect that a statistical-mechanics  model placed on such graphs behaves similarly for any $z>1$. Such an expectation is supported with the exact solution of the Ising model on random graphs~\cite{dorog}, which shows that for any $z>1$, the model has a finite-temperature critical point  belonging to the mean-field universality class. Also the Naming Game exhibits a similar behavior and numerical simulations show that for any $z>1$,  it reaches a global consensus~\cite{lip2017}. However, for directed random graphs, the consensus dynamics does depend on~$z$ and a global consensus appears but only for $z>1.96$ in the Naming Game and for $z>1.85$ in the Ising model~\cite{lip2017}. The present model seems to have a similar behavior with the average node degree~$z$ playing an important role.  A global consensus characterized by nonzero~$m_G$ appears for $z=10$ while for $z=2$ and 3, which is still above the percolation threshold, the model gets trapped in a disordered configuration. A precise location of the transition between these two regimes remains, however, beyond the scope of the present paper.                                                                                                                                                                         

There is a number of models with dynamics driving the system toward consensus, such as, for example, the Voter, Ising, or Naming Game  models. All of them evolve toward consensus but they differ in the details of the evolution. One of the important quantities characterizing their dynamics is a surface tension~\cite{lip2017}, which keeps the interface  (i.e., the boundary between different phases) bounded and is responsible for shrinking droplet excitations. The dynamics of the Ising model or the Naming Game exhibit a number of similarities, such as, for example, a power-law coarsening, which is to a large extent a consequence of the surface tension present in these models~\cite{bray}.   The absence of a surface tension results in a quite different dynamics. Indeed, the Voter model, known to have a tensionless dynamics, exhibits, for example, in the two-dimensional case, a logarithmically slow coarsening and in the three-dimensional version, it does not coarsen at all. Let us notice that in certain disordered systems (spin glasses), the dynamics might also be tensionless~\cite{martin,lipjohnston}. The dynamics of our model on regular networks, which (as shown in Fig~\ref{Fig3}) remains disordered, evolves very slowly and has a well developed interface, may also be tensionless. Possible relations  with some other disordered (and maybe glassy) systems is interesting but beyond the scope of the present  work. As we show in section on population renewal, one can modify the dynamics of our model so that it does not get trapped in a disordered state and most likely evolves as, e.g., an Ising model. It may indicate that in such a way we restore the surface tension into the dynamics.

We do not present here the corresponding numerical results, but we analysed our model also for $N_w>2$ and observed a qualitatively similar behavior. The model gets stucked in a disordered structure for finite-dimensional Cartesian lattices but rather quickly reaches a mono-word phase on a complete graph or sufficiently dense random  graphs.

%%%%%%%%%%%%%%%%%%%%%%%%%%%%%%%%
\subsection*{Multi-object version} \label{multi-object}

\begin{figure}[!h]
\includegraphics[width=\columnwidth]{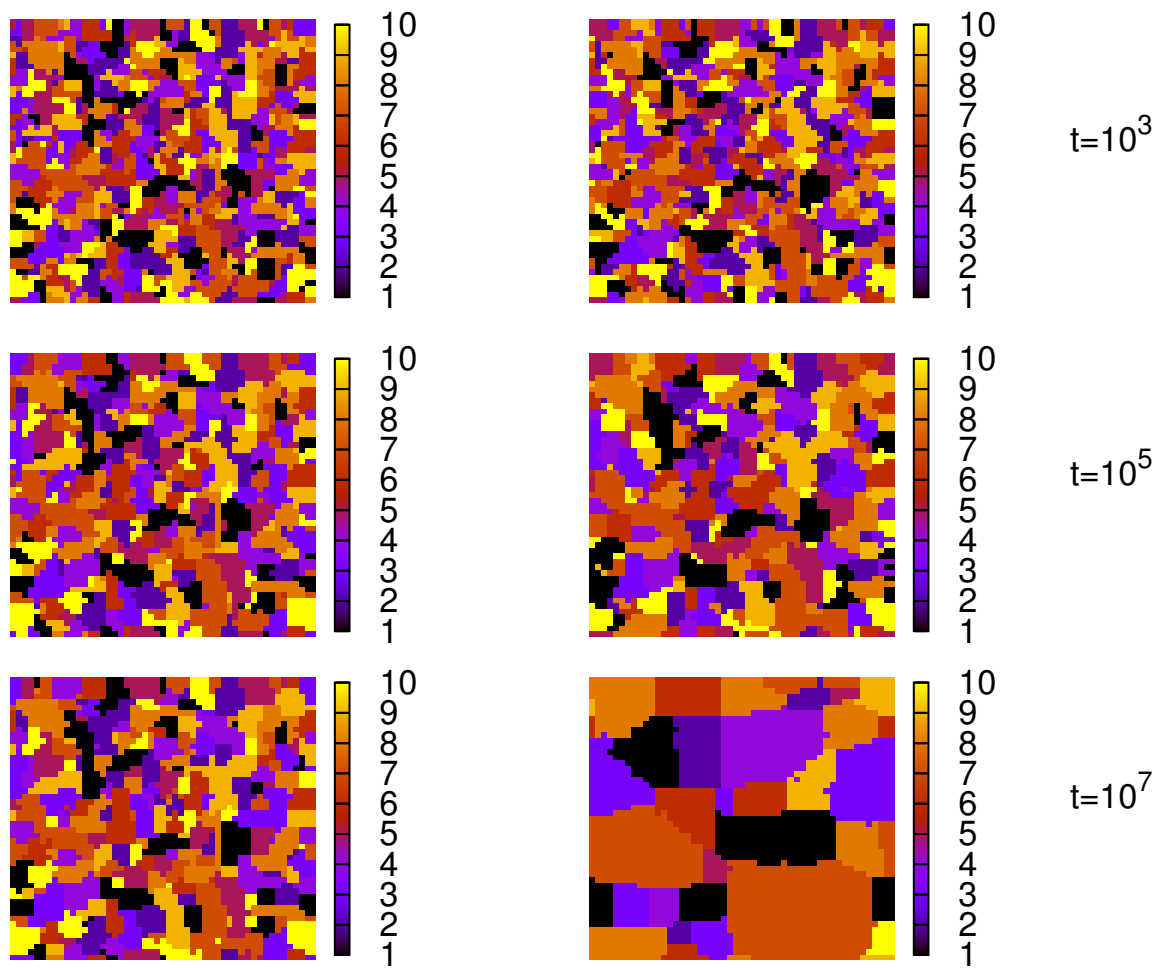}
\caption{ {\bf Distribution of the dominant words that agents use to talk about the first object.}
{\em Left}: simulations on a square lattice  with $N_o=2$, $N_w=10$, $N=50\cdot 50=2.5\cdot 10^3$, \mbox{$\alpha=2$}. {\em Right}: the same simulations but with a population renewal (with probability $p=10^{-5}$).}
\label{Fig7}
\end{figure}

\begin{figure}[]
\begin{center}
\includegraphics[width=13cm]{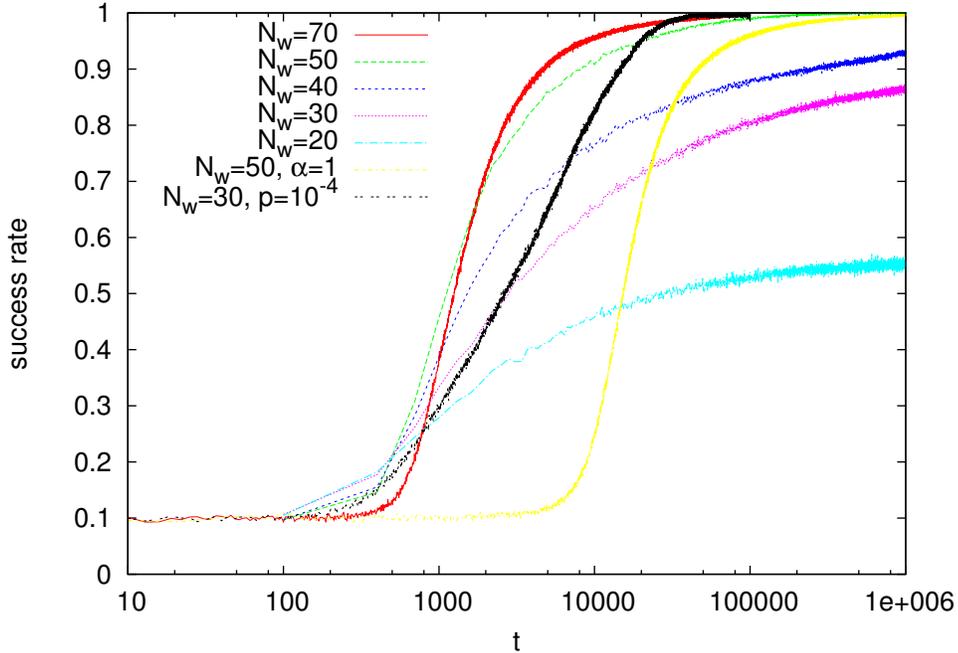}
\end{center}
\vspace{-0.5cm}
\caption{ {\bf Time dependence of the success rate.}
Results for the model on the complete graph of size $N=10^4$ with $N_o=10$, several values of $N_w$, and $\alpha=2$. For $N_w=50$ and $\alpha=1$ (yellow line), we can see a much slower  convergence to a consensus than  for $N_w=50$ and $\alpha=2$. The black line shows the success rate for the version with a population renewal (with probability $p=10^{-4}$).}
\label{Fig8}
\end{figure}

In the previous section, we analysed a model, in which agents try to establish a name for a  single object. 
Here we examine its multi-object generalization. 
On a square lattice, the multi-object version behaves similarly to the single-object one, namely it gets trapped in a disordered configuration, where only some local consensus appears. Indeed, simulations for $N_o=2$ show that only small groups of agents communicate with the same word (Fig~\ref{Fig7}).  We do not present here our numerical results, but a group of agents may reach a fairly good consensus while talking on a certain object, while much worse agreement with respect to another one (in other words, the panels in Fig~\ref{Fig7}, which present the dominant words used by agents for the first object, would be uncorrelated with those for the second object). As in the single-object version, after some initial transient, the evolution of the model nearly stops (in Fig~\ref{Fig7} the configurations for $t=10^5$ and $t=10^7$ are almost identical).
On a complete graph, a much better consensus is reached. During simulations, we measured a success rate defined as a fraction of successful communication attempts in a unit of time. Numerical results for $N_o=10$ show that when $N_w$, i.e., the number of words in agents' inventories, is large enough ($N_w=50$ and 70), the model reaches rather fast a regime, where the success rate is nearly~1 (Fig~\ref{Fig8}). For smaller $N_w$ (20, 30, and 40), the success rate is much lower, even after long simulations. It suggests that large- and small-$N_w$ regimes may be qualitatively different. Let us also notice that for $\alpha=1$,  the regime with the success rate close to unity is reached in time approximately a decade longer than for $\alpha=2$.  Previous simulations in a similar numerical setup but for a much smaller number of agents and shorter time scale suggested that the ordinary reinforcement ($\alpha=1$) does not lead to an optimal communication system~\cite{kirby}.  

%%%%%%%%%%%%%%%%%%%%%%%%%%%%%%%%%%%%%%%%%%%%%%%
%%%%%%%%%%%%%%%%%%%%%%%%%%%%%%%

%%%%%%%%%%%%%%%%%%%%%%%%%%%%%%%%%%%%%%%%%%%%%%%
%%%%%%%%%%%%%%%%%%%%%%%%%%%%%%%

\begin{figure}[!h]
\includegraphics[width=13cm]{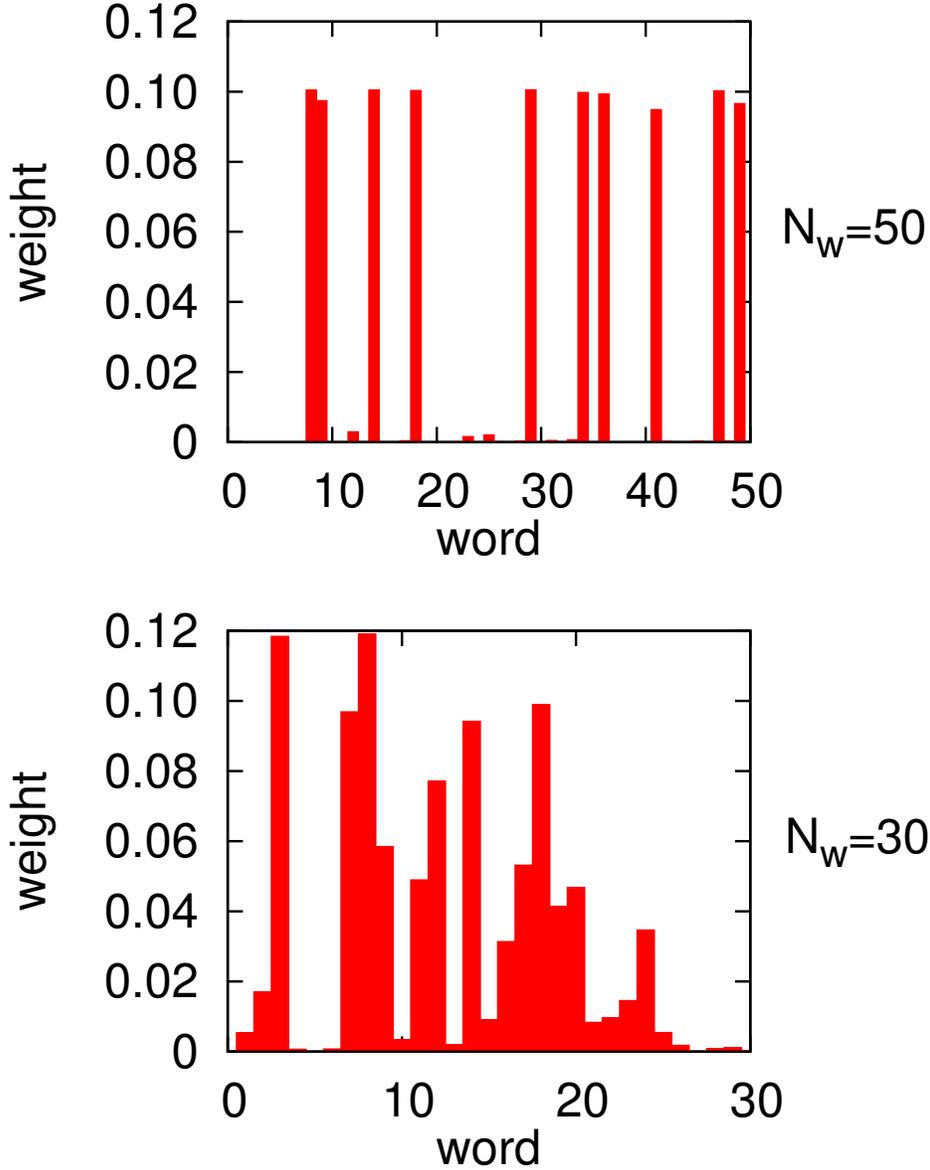}
\vspace{-0.5cm}
\caption{ {\bf Distribution of the total (normalized) weights associated with particular words.}
Results for simulations with $N_o=10$ on the complete graph of size $N=10^4$ and simulation time $t=10^6$ ($\alpha=2$). Simulations for $t=10^5$ lead to nearly identical distributions.}
\label{Fig9}
\end{figure}

Fig~\ref{Fig9} provides yet another indication that the dynamics for large and small $N_w$ considerably differ. In this figure, we present a total weight associated with a given word, defined as follows:
\begin{equation}
w_i^{total}= \sum_{A, j} w_{i,j}(A),   \ \ \ i=1,2,\ldots, N_w
\label{totalweight}
\end{equation}
where summation is over all agents ($A$) and objects ($j$). In Fig~\ref{Fig9}, what is actually plotted is a normalized total weight given as $w_i^{total}/\sum_k w_k^{total}$. For $N_o=10$ and $N_w=50$, we can notice 10~peaks corresponding to the 10~words that are mainly in use. Taking into account a very large success rate (Fig~\ref{Fig8}), it means  that the agents established a single word for each object, which became dominant in their inventories related to this object. As a result only this word is  selected by speakers when they decide to talk about the object and just this word leads then to a correct recognition of the object by hearers. The resulting language provides a nearly perfect one-to-one mapping between objects and words. Let us emphasize that  such a global language emerges spontaneously, as a result of two-agent interactions only. 
%%%%%%%%%%%%%%%%%%%%%%%%%%%%%%%%%%%%%%%%%%%%%%%
%%%%%%%%%%%%%%%%%%%%%%%%%%%%%%%%

Much different behavior can be seen for smaller $N_w$. In Fig~\ref{Fig9}, some peaks can be also distinguished for $N_w=30$, but in addition there is an entire spectrum of less important  but clearly nonnegligible words, which are being used by agents. The one-to-one correspondence between words and objects is missing in this case and the resulting language has a much smaller success rate (Fig~\ref{Fig8}). Since agents increase the weights of the communicated word only when the object is recognized correctly, it means that also nondominant words lead sometimes to a correct recognition---otherwise their  weights (in relation to dominant words) would diminish to zero. It is an analogue of synonymy, a common feature of natural languages. Synonymy, however, does not reduce the success rate, while homonymy (or polysemy~\cite{polysemy}) does. Homonymy appears when a certain word has significant weights associated with more than one object. Communicating such a word may result in an incorrect recognition of the object and the success rate smaller than~1 indicates that homonyms are also present in the emerging  language of our model. The structure of our model is quite complex and some intermediate scenarios are also possible. Namely, with respect to some objects, agents may develop a one-to-one relation between objects and words (like for $N_w=50$), while with respect to some others, a more complex language containing synonyms and homonyms may be used. 

%%%%%%%%%%%%%%%%%%%%%%%%%%%%%%%%
\subsection*{Population renewal} \label{renewal}

As we have already seen, both the single- and multi-object versions of our model on a two-dimensional lattice get trapped in a disordered regime, with only a local consenus reached. In such a regime, the coarsening dynamics, which could lead to formation of larger clusters of agents that reached a consensus, becomes very slow.  Some other models with a consensus dynamics, such as the Ising model or the Naming Game, are known to have much faster coarsening dynamics, which could be attributed to the surface tension generated in these dynamics. 
In this section, we examine our model modified in such a way that the dynamics does not get trapped in a disordered state and induces perhaps some kind of a surface tension. Namely, we introduced a simple mechanism of a population renewal, which means that in each step, the selected agent either (with some probability~$p$)  is replaced with a new one (with all weights reset to~1) or else the agent acts as a speaker. 

\begin{figure}[!h]
\includegraphics[width=\columnwidth]{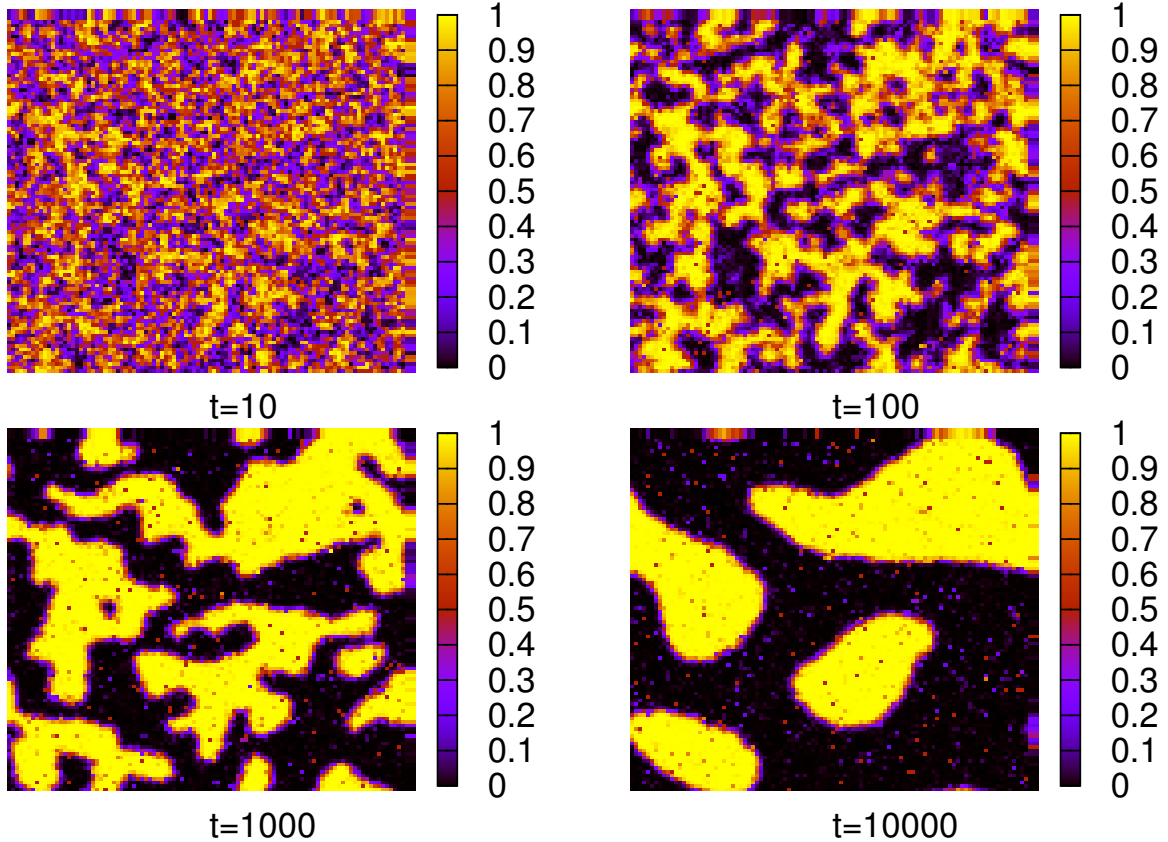}
\caption{ {\bf Spatial distribution of $\boldsymbol {s_1}$, the probability that an agent will select  word $\boldsymbol {W_1}$ (Eq~(\ref{eq1})).}
Results for a single-object model on a square lattice with $N=10^2\cdot 10^2=10^4$, $\alpha=2$, $N_w=2$, and  with the renewal probability $p=10^{-4}$. In this case, contrary to Fig~\ref{Fig3}, clusters of agents with the same dominant word grow steadily.}
\label{Fig10}
\end{figure}

The time evolution of a single-object model with a population renewal on a square lattice is shown in Fig~\ref{Fig10}.  Certainly, the evolution in this case is different than in the absence of population renewal (Fig~\ref{Fig3}). It seems that there is a tendency to reduce the length of interfaces in this model, just as in some models with a surface tension. 

%%%%%%%%%%%%%%%%%%%%%%%%%%%%%%%%%%
%%%%%%%%%%%%%%%%%%%%%%%%%%%%%%%%%%

\begin{figure}[!h]
\begin{center}
\includegraphics[width=13cm]{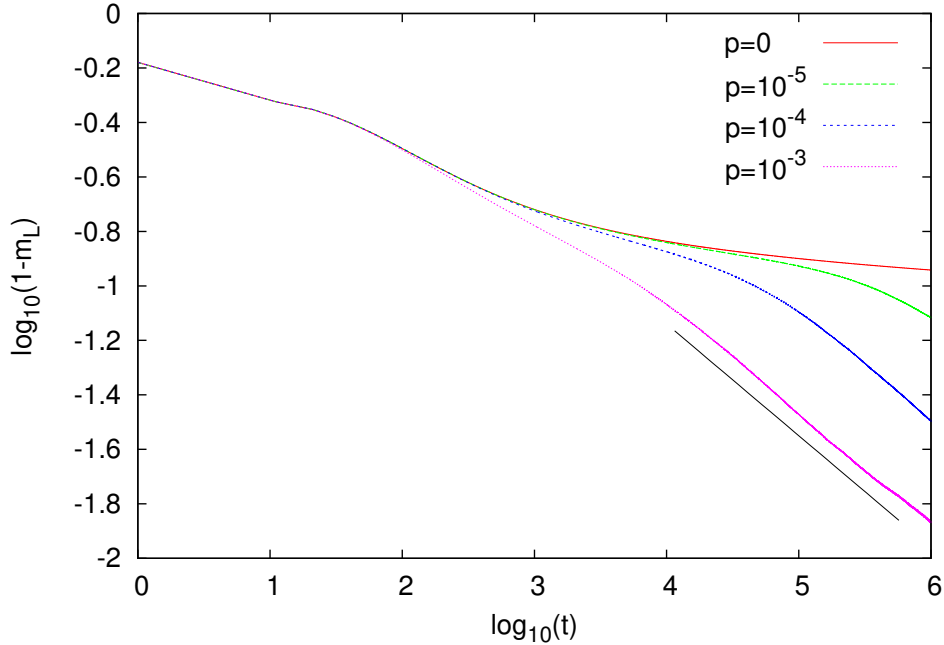}
\end{center}
\caption{ {\bf Time dependence of $\boldsymbol {1-m_L}$ for several values of the renewal probability~$\boldsymbol p$.}
Results for a single-object, square-lattice version of our model. Simulations were made for $\alpha=2$, $N_w=2$, and $N=200\cdot 200=4\cdot 10^4$, and the results are averages of 20 independent runs. The line segment has a slope corresponding to~$t^{-0.41}$.}
\label{Fig11}
\end{figure}

Additional arguments that the dynamics generates some kind of an effective surface tension come from the analysis of time dependence of $1-m_L$ (Fig~\ref{Fig11}). Let us notice that significant contributions to this quantity come mainly from interfacial agents. Provided that the  characteristic cluster size is~$l$, we easily find a scaling relation $1-m_L\sim l^{-1}$~\cite{shore}. From the time dependence of $1-m_L$ (Fig~\ref{Fig11}), we conclude that $l \sim t^{0.41}$ and such a value seems to be independent of the  renewal probability $p>0$. Only for $p=0$, we obtain a much slower increase of~$l$, perhaps logarithmic (in time), which reflects 
the trapping of the model in a disordered configuration as, for example, in Fig~\ref{Fig3}.  A small deviation from the Ising model increase $l\sim t^{1/2}$~\cite{bray} may be attributed perhaps to a diffusive structure of an interface in our model. Let us notice that  a similar  increase $l\sim t^{0.45}$ was observed also in certain opinion-formation models that are expected to have a dynamics with an effective surface-tension~\cite{dallasta}.

%%%%%%%%%%%%%%%%%%%%%%%%%%%%%%%%%%
%%%%%%%%%%%%%%%%%%%%%%%%%%%%%%%%%%

\begin{figure}[!h]
\includegraphics[width=\columnwidth]{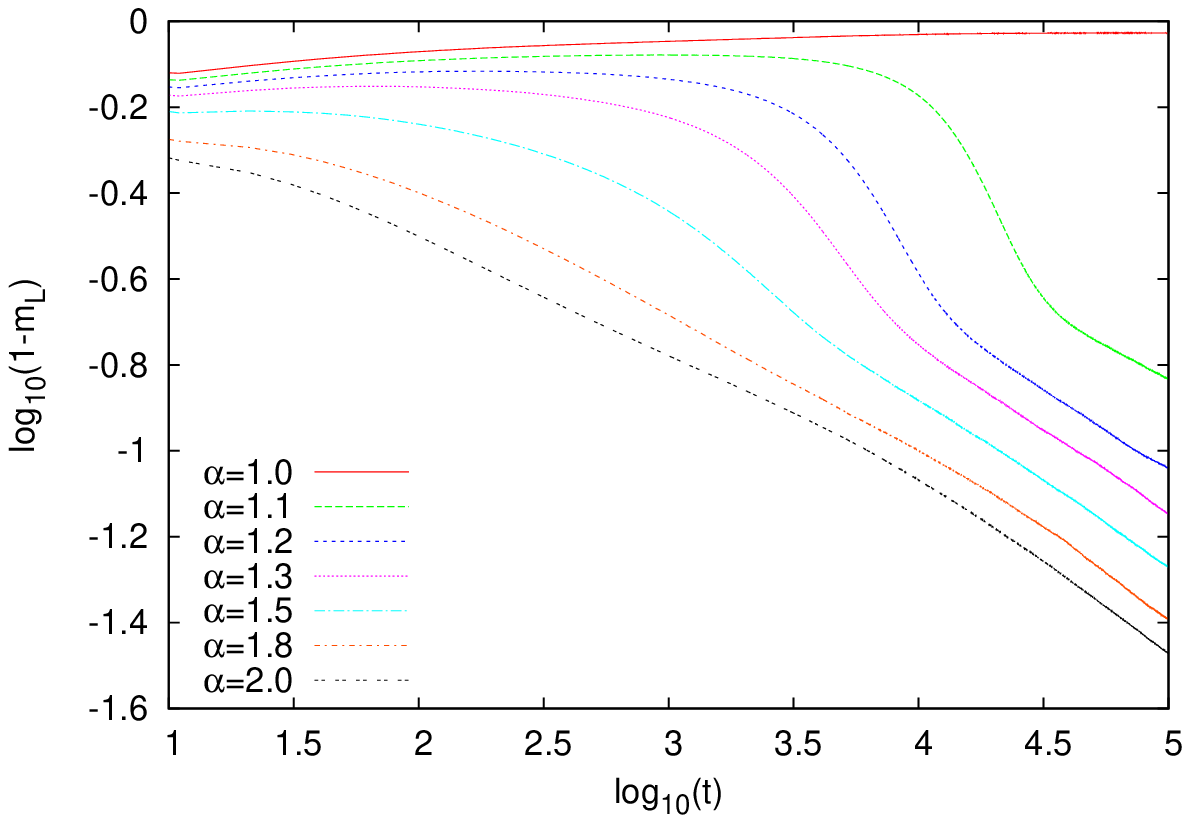}
\caption{ {\bf Time dependence of $\boldsymbol {1-m_L}$ for several values of~$\boldsymbol \alpha$.}
Results for a single-object square-lattice version of our model for the renewal probability $p=10^{-3}$. Simulations were made for $N_w=2$ and $N=200\cdot 200=4\cdot 10^4$, and the results are averages of 20 independent runs.}
\label{Fig12}
\end{figure}

We also analysed how the coarsening dynamics depends on~$\alpha$. Our results for the renewal probability $p=10^{-3}$ are shown in Fig~\ref{Fig12}. It seems that the asymptotic decay of $1-m_L$ is characterized by the same exponent for any $\alpha>1$. A qualitatively different behavior is seen only for $\alpha=1$, where the model does not coarsen. These results suggest that the behavior of our model, also with respect to other properties, should not depend on a particular choice of $\alpha$ as long as $\alpha>1$ (superlinear reinforcement). 

%%%%%%%%%%%%%%%%%%%%%%%%%%%%%%%%%%
%%%%%%%%%%%%%%%%%%%%%%%%%%%%%%%%%%

The population renewal affects also a multi-object version of our model.  In the absence of the renewal, the model on the complete graph (for $N_o=10$ and $N_w=30$) develops a language with a reduced success rate (Fig~\ref{Fig8}) and with no clear object-word mapping (Fig~\ref{Fig9}). However, even a tiny renewal probability ($p=10^{-4}$) considerably increases the success rate of communication (Fig~\ref{Fig8}). We do not present our numerical data here but in this case, a clear object-word mapping does emerge, similar to that in the upper panel of Fig~\ref{Fig9}. Also a multi-object square-lattice version of the model  behaves differently for $p>0$. Indeed, the snapshot configurations (Fig~\ref{Fig7}) show  that in this case the model does not get trapped (as for $p=0$) but coarsens similarly to the single-object version with population renewal (Fig~\ref{Fig10}). 

%%%%%%%%%%%%%%%%%%%%%%%%%%%%%%%%%%%%
%%%%%%%%%%%%%%%%%%%%%%%%%%%%%%%%%%%%

The overall behavior of our models is summarized in Table \ref{table1}.
\begin{table}[h]
\caption{Behavior of our models as a function of dynamics and network structure}
\begin{center}
\begin{tabular}{|c|c| p{5cm}|} \hline
\textbf {Dynamics} & \textbf {Network} & \textbf {\hfil Behavior} \\  \hline
\multicolumn{3}{|c|}{\textbf{Single-Object}}\\ \hline
 & Cartesian & local consensus \\ \cline{2-3}
{$\alpha>1$, no renewal} & Complete graph & global consensus \\ \cline{2-3}
 & Random Graph & global/local consensus  for large/small $z$ \\ \hline
{$\alpha>1$,  renewal} & Cartesian & global consensus, power-law coarsening \\ \hline
 & Cartesian & no local consensus \\ \cline{2-3}
{$\alpha=1$, no renewal} & Complete graph & no local consensus \\ \hline
{$\alpha=1$, renewal} & Cartesian & no local consensus \\  \hline
\multicolumn{3}{|c|}{\textbf{Multi-Object}}\\ \hline
{$\alpha>1$,  no renewal} & Cartesian & partial local consensus \\ \hline
{$\alpha>1$,  no renewal, large $N_w$} & Complete graph & unique object-word mapping \\ \hline
{$\alpha>1$,  no renewal, small $N_w$} & Complete graph & language with homonyms and synonyms \\ \hline
{$\alpha>1$,  renewal, small $N_w$} & Complete graph & unique object-word mapping \\ \hline
{$\alpha>1$,  renewal} & Cartesian & unique object-word mapping, power-law coarsening \\ \hline
\end{tabular}
\end{center}

\vspace{3mm}
\caption* {Global consensus means that all agents communicate using the same word. Local consensus means that only neighbouring agents use the same word. Partial local consensus means that neighbouring agents use the same words to communicate only with respect to some objects while with respect to others, there may be no efficient communication.} 
\label{table1}
\end{table}
\section*{Discussion and conclusions}

The objective of the present study was to examine the emergence of linguistic
conventions in a multi-agent model with reinforcement learning. Models of this kind
have a potential to generate a complex language, which reflects their multi-object and
multi-agent structure, but global aspects of their dynamics are rather purely
understood. This is much in contrast to some simpler models with agreement dynamics,
like Ising model or Naming Game, which due to the surface tension have curvature-driven
dynamics \cite{bray}, or voter model, which lacks the surface tension and has a much different
dynamics \cite{krapivsky}.

In the single-object version, agents do not need to recognize the object and each communication attempt is in a sense successful. It turns out that it is the structure of the network of interactions between agents that plays a decisive role and determines the asymptotic state of the model. While for a complete graph or sufficiently dense random graphs, a global consensus is reached and all agents use the same word for communication, on finite-dimensional lattices or sparse random graphs, only a local consensus is reached and the model gets trapped in a disordered configuration.

In the Naming Game, which is an alternative model describing formation of linguistic consensus, agents also try to establish a name for a given object. However, in this case a global consensus is much easier to reach except on networks with a strong community structure~\cite{barrat2008,lipadaptive}. Such a strong tendency to reach a consensus could be explained using the notion of an effective surface tension, which is generated in the Naming Game. It turns out that when a population renewal is introduced, the surface tension emerges also in our model and its evolution toward consensus is much enhanced. However, such a curvature-driven evolution appears only for $\alpha >1$, which indicates the importance of a superlinear reinforcement.  Our study thus shows that a physical intuition developed for some statistical-mechanics models may be also used to understand to some extent reinforcement learning systems.

Having in mind linguistic contexts, the multi-object version of our model is more interesting. The results show that in this case the structure of the network plays an important role as well. On complete graphs, an efficient global communication may be established, such that all agents unambiguously match each object with the same corresponding word. On finite-dimensional lattices, such a mapping is again only local (and partial) and the model gets trapped in a disordered configuration. In the multi-object version, in addition to the network structure, other parameters are also important, namely the number of objects~$N_o$ and the number of words that agents have at their disposal~$N_w$. Our simulations suggest that a unique object-word matching may emerge only when $N_w$ is considerably greater than~$N_o$. If this is not the case, the resulting communication is less efficient and the emerging language contains some homonyms and synonyms. Of course, such behavior should be by no means considered as undesirable or unrealistic, since all natural languages contain such forms. Further studies concerning, for example, the frequency and durability of homonyms and synonyms in our model would be desirable, but are left for the future. 

In the multi-object version, the population renewal also enhances formation of an efficient communication. The population renewal basically resets the weights of an agent, and thus it plays a role similar to forgetting, which is a factor already known to improve the performance of reinforcement learning systems~\cite{barrettzollman}. Our snapshot configurations show that also in this case, the population renewal most likely induces a certain surface tension, similarly to the single-object version. Hence, one of the merits of our work is the demonstration that reinforcement learning systems with the population renewal and superlinear reinforcement ($\alpha>1$) reveal certain similarities to some other models with the agreement dynamics (such as the Naming Game) and exhibit a power-law coarsening. However, without (or perhaps with a sufficiently small) population renewal or for (sub-)linear reinforcement ($\alpha\leq 1$), the dynamics of these two systems considerably differ.

Let us notice that the surface tension might be of some importance also in linguistic processes and, for example, some recent works show that the boundaries of dialect  regions are controlled by a length-minimizing effect analogous to the surface tension~\cite{human-dialects}. Moreover, the fast extinction of natural languages, especially those of a small number of users, indicates that some coarsening does take place. Hopefully, some simple models can be propounded, which might provide some insight into such a linguistic dynamics. Of course, the processes of emergence, diversification or extinction of languages are very complex and affected  by a large number of factors such as, for example, politics, geography, economy, or technological development. Thus computational modeling may provide their very crude, qualitative description at most.

Population renewal supplies a new generation of language users. Similarly as children, they quickly learn a language of the neighbors they interact with. Let us notice that some linguists strongly advocate the view that profound language changes occur in the process of language learning and children perhaps play an important role in this process, for example, making mistakes~ \cite{lightfoot1999,lightfoot2006}. However, such a view can be questioned  because the modifications children generate seldom survive till their adulthood~\cite{kerswill,holger} due to, for example, usually lower social and economic status of youngsters~\cite{labov2001}. In our opinion, it would be certainly interesting, as well as feasible, to consider an aged-structured version of our model and analyse the role of the young generation. As we have already noticed, young users are needed to generate a surface tension and coarsening (which in the context of human language evolution is probably more realistic than a population  trapped in a multilanguage regime). In the aged-structured population, a peer communication is an expected feature, but with such preference being too strong, the population could split up into separate linguistic communities. Analysing the emergence of a young generation dialect and its possible influence on the language of adults is, however, left as a future problem.

%\nolinenumbers

\end{document}